# Query Complexity Based Optimal Processing of Raw Data

Mayank Patel
*Distributed Databases Group*
*DA-IICT*
Gandhinagar, Gujarat, India
mayank@daiict.ac.in

Minal Bhise
*Distributed Databases Group*
*DA-IICT*
Gandhinagar, Gujarat, India
minal_bhise@daiict.ac.in

***Abstract*— The paper aims to find an efficient way for processing large datasets having different types of workload queries with minimal replication. The work first identifies the complexity of queries best suited for the given data processing tool. The paper proposes Query Complexity Aware partitioning technique QCA with a lightweight query identification and partitioning algorithm. Different replication approaches have been studied to cover more use-cases for different application workloads. The technique is demonstrated using a scientific dataset known as Sloan Digital Sky Survey SDSS. The results show workload execution time WET reduced by 94.6% using only 6.7% of the dataset in loaded format compared to the original dataset. The QCA technique also reduced multi-node replication by 5.8x times compared to state-of-the-art workload aware WA techniques. The multi-node and multi-core execution of workload using QCA proposed partitions reduced WET by 42.66% and 25.46% compared to WA.**

## I. Introduction

Modern applications and scientific datasets sizes are increasing rapidly. The use of smart devices and high-frequency sensors generates enormous amounts of data. The Large Hadron Collider LHC experiments at CERN generated 3.6PB of data every year during its initial runs in 2009, which increased to 90PB by 2018 [1]. The Sloan Digital Sky Survey SDSS has generated 245TB of new data in its latest data release DR-17 [2]. The DR-17 size is 87.5times compared to its DR-1, released in 2003. Fig.1 shows scientific experiment dataset growth like NASA's Earth Observing System Data and Information System (EOSDIS) [2]. It can be observed that the EOSDIS dataset will double in size in just one year from 2021 to 2022. The efficient access of such datasets is challenging.

The deployment of such applications requires a distributed environment to process large datasets in parallel. The EOSDIS dataset is also migrated to a commercial cloud environment [2]. The cloud service providers need to retain copies of the stored data to handle hardware failures and provide faster access to the data around the globe. It is known that modern applications have different query processing requirements. One database management system DBMS cannot handle all kinds of query types efficiently. For example, row stores are designed to process transactional queries, while column stores are best suited for analytical queries. Therefore, real-world applications are using multiple database management systems known as Hybrid Transactional and Analytical Processing HTAP systems. The issue with such systems is that they need to load data into their private databases, increasing replication and data loading time [3]. The raw or in-situ engines work directly on raw data to eliminate high data loading time and replication issues. However, in-situ engines are not optimized to execute all types of workload queries and require more resources due to data reparsing issues. Raw data needs to be loaded into a DBMS to execute complex and frequent queries faster.

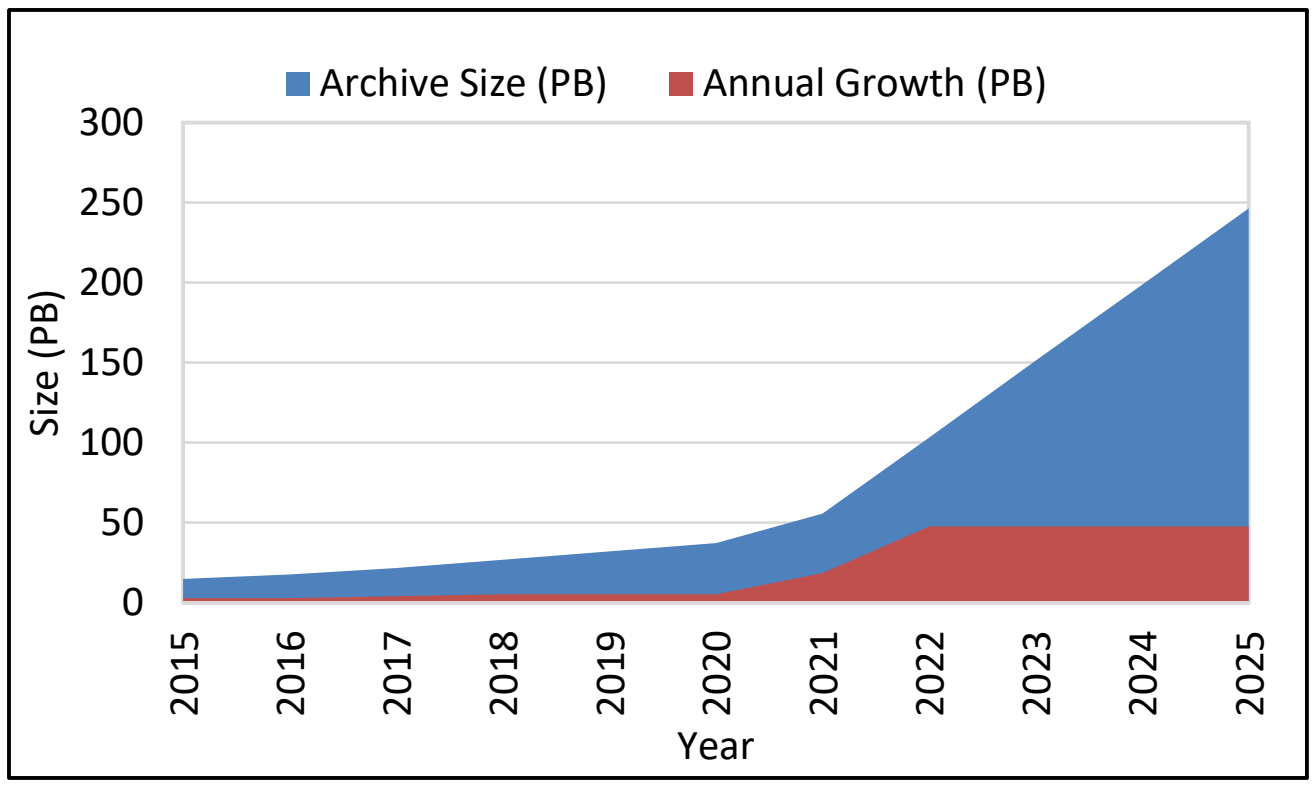

Fig. 1. Scientific Experiment Dataset Growth [2]

### A. *Motivation*

The size of application data is growing with time. HTAP systems need to replicate the entire dataset into multiple DBMS systems. In-situ engines allow query processing on raw data without data loading or replication. In-situ or raw engines can provide results of several queries before data loading steps get completed by traditional DBMS [4], [5]. Replication & distribution of data among raw and database formats is scarcely studied. In addition, existing cost function based solutions are complex and time-consuming.

### B. *Problem statement:*

Most data management tools are optimized to perform a few types of tasks. For example, row store DBMSs are optimized to handle complex transactional queries efficiently, while column stores execute analytical queries faster. In comparison, in-situ engines eliminate data loading time but incur high QET time. Advance in-situ engines cache the processed data into main memory to reduce QET time for future queries. The first challenge is identifying which tool is best for which query types. Another challenge is to find an efficient way of distributing raw data among multiple tools, which reduces total workload execution time WET.

### C. Paper Contributions

- The paper identified that raw engines could efficiently handle simple zero-join queries while complex multi-join queries are faster in DBMS.
- The work distributed workload in simple and complex query types using a lightweight query complexity identification logic.
- Paper proposes a lightweight query complexity aware partitioning algorithm QCA to partition the data among raw files and database.
- Multiple replication strategies have been explored to reduce the data replication and WET.
- Result analysis of single-core and multi-core parallel processing experiments shows the benefits of using proposed QCA technique on modern multi-core CPUs and distributed environments.

## II. Related Work

Related work discusses different approaches used by researchers to process raw data with replication strategies.

The in-situ processing does not require raw data to be loaded into a specific database format. Initial in-situ engines accessed actual raw files to query the data to provide results and discarded processed data as soon as results were generated [6], [7]. It raised reparsing issues which utilized resources multiple times to process the same raw data when needed by other queries. Researchers proposed caching and indexing of processed raw data into main memory to moderate reparsing issue while achieving zero data replication [5], [8], [9]. However, processed data needs to be cleared from the main memory to accommodate new data. Researchers proposed to load the parsed data into row store or column store DBMS to eliminate the reparsing issue [10]–[12]. The incremental loading of data into DBMS reduced upfront loading time. However, these techniques may end up loading the entire dataset into DBMS, which is a replica of the raw data into DBMS format.

Researchers have proposed limiting the amount of data to be loaded into DBMS by checking the cost of data access from raw and database formats [13]. The techniques proposed to load fix number of attributes into databases based on a given storage budget. Another technique improved the storage utilization by 5% by considering actual storage requirements of attributes that can be loaded in a given storage budget [14]. Both techniques load a small partition of the entire dataset accessed by workload queries reducing replication. However, none of these techniques consider replication of data into raw format and database format in a multi-node distributed environment. Ample research papers exist that discuss the distribution of single format data among multiple nodes using full replication or partial replication of dataset to reduce internode communications [15]–[17]. Researchers have proposed to change vertical partitions dynamically based on workload analysis to reduce query execution time while minimizing internode communication and replication [15]. Researchers have also proposed adaptive partial replication technique, which uses dictionary mapping of frequently accessed attributes [17]. The attributes with more frequency than a threshold value are replicated among multiple nodes. The threshold value gets modified to incorporate workload changes. Other techniques like PDC use horizontal partitioning and distribution of data among multiple nodes [18]. The PDC technique executes the query on required partitions distributed among multiple nodes in parallel and combines results received from those nodes to produce the final result.

Most techniques discussed in this section suffer from high data to result time because they require data loaded in databases or spend more time calculating costs to partition the dataset. The paper proposes a lightweight partitioning technique that classifies queries and distributes raw datasets among multiple nodes in raw and database formats to achieve optimal workload execution time.

## III. Query complexity aware Partitioning Technique-QCA

This section discusses data structure and algorithm logic for the Query Complexity Aware QCA partitioning technique.

### A. Data structures

QCA algorithm uses simple lists and key-value dictionaries to store the query attributes, query complexity, and relevant partitions. QCA uses schema dictionary, workload list, workload attributes, and query dictionaries generated using schema and workload extraction functions similar to WSAC [14]. Table 1 shows workload list *w_l* populated by reading workload file. The *QT* dictionary stores query types identified by the query complexity identification steps. *QT* values are shown in table 2. The value 0 means the query type is simple, and 1 for the complex type. *QT_P0* and *QT_P1* store list of attributes coming in simple and complex queries. Common attributes for both partitions are stored in the common attributes partition *CAP* list. *PC_Q0* and *PC_Q1* list keep track of partially covered queries based on *CAP* list.

Table I. Workload list W_L

| 0[T_ID] | 1[Statement] |
|---|---|
| TRUN | “TRUNCATE TABLE PhotoPrimary;” |
| COPY | “COPY PhotoPrimary FROM '/…SDSS/PhotoPrimary.csv' (DELIMITER ',');” |
| Q0 | “Select count(objID) from PhotoPrimary;” |
| Q4 | “SELECT objID, ra ,dec FROM PhotoPrimary WHERE ra > 185 and ra< 185.1 AND dec > 56.2 and dec < 56.3 limit 100;” |

Table II. Query Type Dictionary QT

| Key (Q_ID) | 1 | 2 | 3 | 4 | 5 | 6 | 7 | 9 | 10 | 11 | 12 |
|---|---|---|---|---|---|---|---|---|---|---|---|
| Value(Query Type) | 1 | 0 | 1 | 0 | 1 | 0 | 0 | 1 | 0 | 1 | 1 |

### B. QCA Algorithm

The idea behind the QCA technique is to partition the dataset and distribute the workload in such a way that queries performing faster on a given tool can be allocated to that tool. On the contrary to the HTAP systems, QCA tries to achieve faster query execution times with minimal replication and loaded data partitions.

```
Algorithm 1: QCA Partitioning

   Data: w_l   = Workload List
          QT   = Query Types Dictionary
         q_l   = Query List
       que_d   = Dictionary of Queries
         s_d   = Schema Dictionary
   Result: SQ-Raw, CQ-DB & CAP Partitions

1.  QCI(w_l, que_d, s_d):
2.  #Query Complexity Identification
3.  For each task T in w_l do
4.    If T.Statement has multiple tables
5.          QT[T.Q_ID] = 1
6.    Else
7.          QT[T.Q_ID] = 0
8.  End
9.  Get QT_P0, QT_P1 = GRA(que_d, QT)
10. CAP = QT_P0 ∩ QT_P1 #Common Attributes
11. QT2 = PCQ(que_d, QT_P0-CAP, w_l)
12. QT3 = PCQ(que_d, QT_P1-CAP, w_l)
13. Repeat steps 8 to 12 for QT2, QT3.

14. Function GRA(que_d, QT)
15. #Grouping of Attributes
16. For each query i in que_d:
17.   For each attribute j in que_d[i]:
18.        If QT[i] == 0
19.             Add j in QT_P0
20.        Else
21.             Add j in QT_P1
22.   End
23. End
24. Return QT_P0, QT_P1

25. Function PCQ(que_d, QT_P, w_l)
26. #Identification: Part. covered Queries
27. QT = [0]*w_l.length;#Assign 0s
28. For each query i in que_d:
29.   For each attribute j in que_d[i]:
30.        If j not in QT_P list
31.             QT[i] = 1 #New QT list
32.   End
33. End
34. Return QT;
```

The proposed workload and query complexity aware algorithm uses lightweight query identification and partitioning steps to reduce algorithm execution time AET compared to other cost based techniques [13], [14]. QCA algorithm can be divided into three parts, 1) Query Complexity Identification QCI, 2) Grouping of Attributes based on query classification GRA, & 3) Identification of Partially covered Queries PCQ. QCA algorithm first identifies the type of queries best suited for a given tool using initial results. The initial experiment results have been plotted in fig. 3. The analysis had shown that zero-join queries were performing faster in raw engines than traditional DBMS. While queries having multiple joins were slow in raw engines. Therefore, the technique classified the query workload into two query types. The first type is for simple queries SQ, which contained zero join. The second query type included remaining one or more join queries. This second category of queries is called complex queries CQ in this paper.

The proposed QCA technique uses general attribute and table name extraction functions from schema and query workload to populate workload list, schema dictionary, and query dictionary [14]. The QCA algorithm first identifies the SQ and CQ type queries stored in the workload list. The algorithm uses the simple logic of counting no. of tables present in the query statement. If two or more table instances are found in the query statement, then the query is classified as a complex query. The technique also considers self-join queries as complex queries. Table-II shows the query ID and query complexity type updated in *QT* as key-value pair. The single table instance queries are classified as simple queries SQ. The GRA function groups the SQ and CQ attributes in two different lists *QT_P0* and *QT_P1*. The intersection of these two lists provides the list of common attributes partition *CAP*. The SQ partition is *QT_P0,* and the CQ partition is *QT_P1* after the first round of QCA partitioning. The SQ partition can be stored in raw format, while the CQ partition needs to be loaded in DBMS.

The QCA algorithm can further refine the partitions based on output from partially covered queries *PCQ* list as new query type *QT* input. Steps 8-12 can be repeated until all workload queries get covered in *QT2* union *QT3* to further partition raw and database partitions. These steps reduce partition size and find new group of queries covered by smaller partitions. The partition refinement benefits broad table datasets with a large number of distinct queries in the workload. For most cases, the first round of partitioning might be enough to partition a single table into the raw format for SQ queries *QT_P0* and database format for CQ queries *QT_P1* where *CAP* gets replicated in both partitions. The following section discusses data distribution cases for *CAP* partition.

### *C. Data Distribution Cases*

All the partitions received from the QCA technique having zero common attributes can be kept in their respective raw or database formats. For example, if a *QT_P0* raw partition covers 4 out of 5 simple queries with no attributes of the fifth query, it can be kept in raw format. This can also be interpreted as all attributes required by the fifth query would be in one or more common attribute partitions *CAP*s. For simplicity, consider that there is only one *CAP* partition. The *CAP* and remaining partitions can be arranged in five different ways. This section discusses all five cases with their advantage and disadvantages based on the location of *CAP*.

*1) No Replication cases*

The cases discussed in this section do not replicate common attributes among raw or database formats. Therefore, a data processing tool capable of executing join queries on multi-format data is required.

*a) Case-1:* Keep the *QT_P1* which includes *CAP* in the loaded format. The remaining partition *QT_P0-CAP* can be stored in raw format. This ensures that complex queries have the best query response times. However, simple queries requiring data from *CAP* will have to perform join with database partition *QT_P1*, which might increase QET for partially covered queries.

*b) Case-II:* Keep the *QT_P0* which includes *CAP* in the raw format. This arrangement allows simple queries to execute without any joins. However, complex queries will require joining *QT_P1-CAP* partition stored in database format with raw partition *QT_P0*. This case has minimal attributes in the loaded format. Therefore, CQ queries may suffer from high QET.

*c) Case-III & IV:* In this arrangement, the *QT_P0-CAP* stays as a raw partition, and the *QT_P1-CAP* partition is loaded in the database. The difference between case-III and case-IV is the location of the *CAP* partition. The *CAP* partition can be either loaded in the database or kept in raw format. The benefit of this case is that all fully covered queries that do not require the *CAP* partition may execute a little faster due to the smaller partition size. However, all partially covered queries PCQ have to be joined with *CAP* partition. The additional JOIN operation may increase WET time due to the high number of PCQ queries in case-III & IV compared to other cases.

*2) Replication case*

The HTAP systems and most query workload balancing techniques require replication of data on all nodes. QCA technique limits the replication to only *CAP* partitions. Replication of *CAP* with raw and database partitions provides freedom to choose different tools for different format partitions which can not join different format partitions.

*a) Case-V:* The *QT_P0* and *QT_P1* partitions that cover all simple and complex queries have to be used in this case. Both partitions include common attributes CAP, which allows the execution of queries using a single partition with no additional joins. This case also eliminates internode communication in a distributed setup.

## IV. Experimental Setup

This section describes the experimental setup consisting of Hardware & Software Setup, Dataset & Query Set.

### A. Hardware & Software Setup

The machine used for experiments has a quad-core Intel i5-6500 CPU. The maximum CPU speed is 3.20GHz. RAM capacity is 16GB with in-built Intel HD Graphics 530. The machine is running a 64-bit Linux operating system Ubuntu 18.04 LTS. For permanent storage, a SATA hard disk drive having 500GB of storage space is used to store raw files and databases. The disk rotation speed is 7200 rotations per minute. The QCA algorithm is developed using python language using Jupyter Notebook. The raw data query processing framework is used to query the raw data directly [4]. The framework can also load data into PostgreSQL to execute complex queries. The *top* and *iotop* tools were used to monitor CPU, RAM & IO resource utilization.

### B. Dataset & Query Set

A real-world scientific dataset and query set has been used to apply the QCA algorithm. The dataset is an astronomy data dataset called Sloan Digital Sky Survey SDSS. It is a major multi-spectral imaging and spectroscopic redshift survey. A dedicated 2.5-m wide-angle optical telescope is used to observe the sky. The telescope is located at Apache Point Observatory in New Mexico, United States. The DR-16 data release having 118TB of new data is used for experiments [19]. There are 134 tables and 59 views in the data release. The SDSS keeps track of queries executed on the dataset and logs them in *SQLlogAll* table. Analysis of the query workload of 0.4M showed that 55% of queries of the workload belong to a single view named *PhotoPrimary* view. Therefore, 4M records from *PhotoPrimary* view were extracted for the experiment purpose to cover maximum workload queries. The top 1000 unique queries executed on *PhotoPrimary* view were extracted from recorded 0.4M DR-16 queries. These queries represented 51% workload of the DR-16 at the time of access. The workload of 12 queries has been created based on the similarity of the attributes & query types.

### C. Experiment Flow

We have performed multiple experiments to evaluate the proposed QCA technique. Initial experiments have been done with the original dataset to find actual WET using state-of-the-art DBMS PostgreSQL and In-situ engine PostgresRAW [5]. Section V-A shows the results achieved by executing workload on the original dataset.

*1) QCA Technique*

Multiple experiments have been performed for QCA proposed partitions to evaluate the performance of QCA in typical single-core execution and resource optimization modes.

*a) Single Core Execution*

Fig. 2 shows the experimental flow of applying QCA technique to the SDSS dataset. The QCA technique takes dataset schema, workload list, and actual raw files of datasets in CSV format as input. The extract from schema function uses DDL files of schema to find all table and their attributes of a dataset and stores them in *s_d* dictionary. The extract from workload function creates a dictionary of workload queries *que_d* containing table names and attributes used in a statement based on *s_d*. The QCA algorithm first identifies the types of queries by checking the number of tables coming in a query. Once the query complexity identification is complete, the attributes of similar types of queries are grouped to form simple query partition *QT_P0* to be stored as raw and complex query partition

*QT_P1* that needs to be loaded in the database. The intersection of both the list provides *CAP* list. The first iteration of QCA ended with all queries partially covered because all the workload queries had 2 or more attributes in *CAP*. Out of 54 workload attributes, database partition had 25 attributes, 21 in raw partition and 10 attributes in *CAP* with primary key attributes replicated in all partitions to allow JOINs. The different sets of partitions have been created to cover three out of five cases discussed in section III-C.

The experiment section V-B discusses results obtained for data distribution cases I, II, and V. The cases III & IV were not included because all queries would require JOIN, which would be slower than other cases where 41-58% of workload queries did not require any additional JOINs. The different QCA case results have been compared with the original dataset execution on PostgresRAW[5] and workload aware WA partitioning techniques by Zhao[13] and WSAC[14]. The best result of the WA techniques is considered, where all the required data is loaded in DBMS due to availability of enough storage budget.

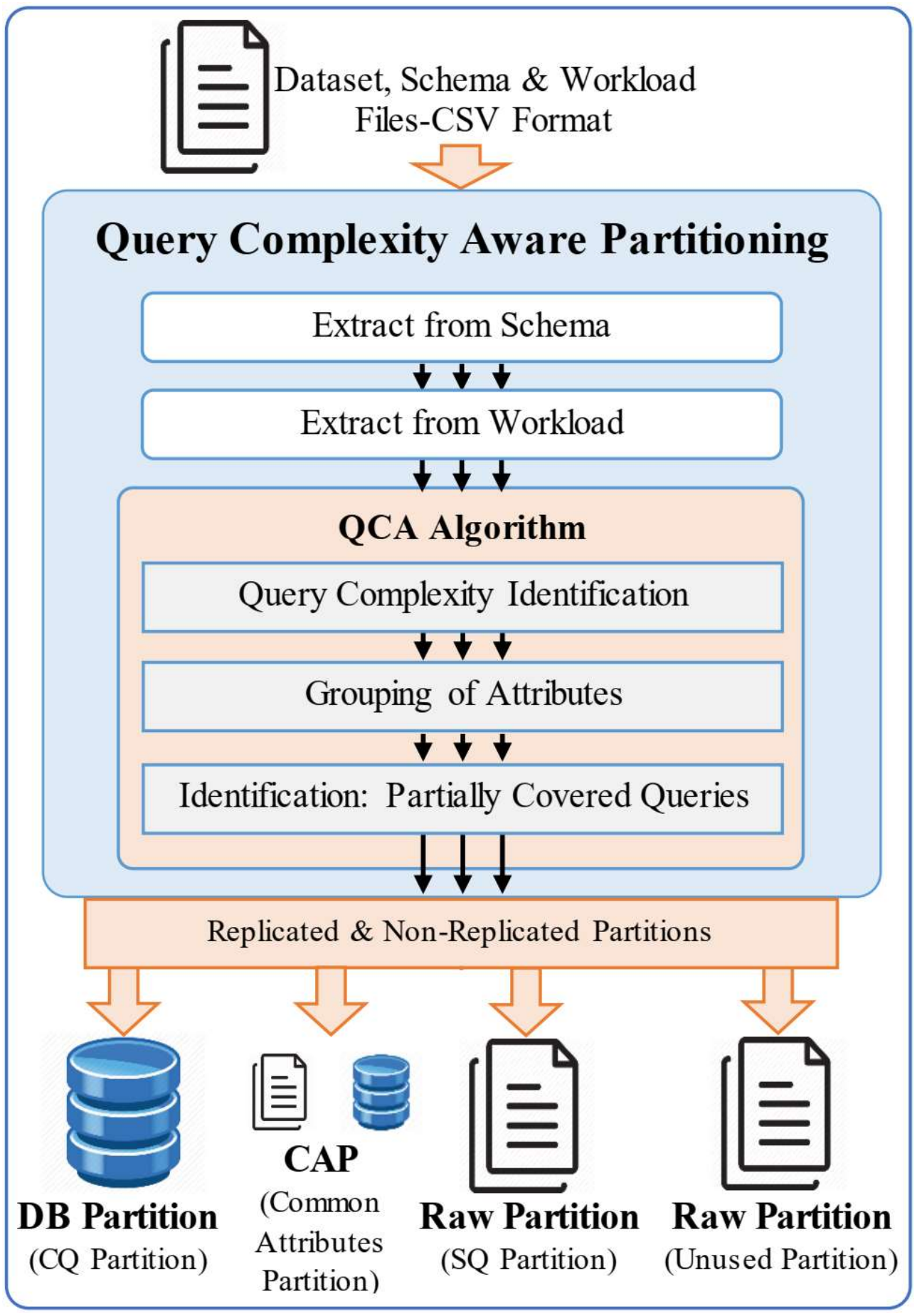

Fig. 2. Experiment Flow: QCA Technique

*b) Optimizing Resource Utilization*

The resource utilization analysis for initial experiments had shown that CPU, RAM, and IO resources stay underutilized. It is known that parallel loading on disk storage devices does not improve DLT time [20]. The parallel loading on a multi-node setup would require distributing a single query to multiple nodes and combining partial results to achieve the final result [18]. The benefit of the QCA technique is the execution of simple queries on raw data, which reduces the DLT time of workload. More CPU time can be utilized to execute queries in parallel as the DLT time gets reduced. The QCA resource optimization experiments have been setup to benefit from these observations to utilize available resources and improve WET.

For the first experiment, let us assume a 2-node setup build to handle the SDSS workload. The workload aware techniques would have to replicate the hot partition on both nodes to redirect queries on an underutilized node to balance the load. A best-case is considered where a load balancing technique has distributed workload equally on both nodes. On the other hand, the QCA technique needs to load only the complex query CQ partition on a single node to cover 7 out of 12 queries. At the same time, the remaining 5 simple queries can be executed on the second node without loading any data.

The second experiment tries to utilize available resources for efficient execution of workload on a single node for a modern quad-core CPU. The WA techniques have to wait till the loading of the hot partition completes using a single CPU core. After that, query execution tasks have been executed in parallel to utilize all available CPU cores. The QCA technique requires loading CQ partition using a single CPU core. While simple queries have been executed in parallel, as PostgresRAW does not require IO once all data is cached in memory. The five simple queries can be executed in parallel with each other, but PostgresRAW keeps the cached data separate, which would require all queries to fetch data from raw files for each connection. Therefore, the execution on simple queries was done in sequence using a single connection to benefit from data cached by the first query and optimize CPU utilization. The SQ queries completed execution in parallel before CQ partition loading got completed. Therefore, once loading of the CQ partition was completed, all the CQ queries were scheduled to execute in parallel using all available CPU cores. Results have been discussed in *Optimizing Resource Utilization* section V-B.

## V. RESULTS & DISCUSSION

This section discusses the experiment results performed using the SDSS dataset having 4.1GB size and 1M records. For convenience, the PgSQL and PgRAW aliases have been used to represent PostgreSQL and NoDB implementation PostgresRAW [5]. The result section includes the original dataset WET before and after applying QCA technique. The resource utilization comparison is also added to understand resource cost savings. Finally, the multi-node and parallel processing scenarios have been discussed to show the advantages of the proposed QCA technique.

### *A. Original Dataset WET*

Fig. 3 shows the comparison of WET for SDSS dataset on state-of-the-art row store DBMS PostgreSQL with NoDB raw engine PostgresRAW. It can be seen that PgSQL suffers from high data loading time DLT. However, simple and complex query execution time QET is 4x to 42x times better than PgRAW. Fig. 4 shows the QET of simple queries in PgSQL and PgRAW. This detailed comparison of simple queries disclosed that PgRAW outperforms PgSQL in 4 out of 5 queries. It can be seen that initial query Q2 is slow in PgRAW because it needs to process attributes from raw files. All the data gets cached in the main memory by the time sequential workload execution reaches Q4. Therefore, all remaining simple queries are faster than PgSQL. QCA technique proposes a way to identify simple queries and tries to improve QET by accessing only required partitions for the respective tools.

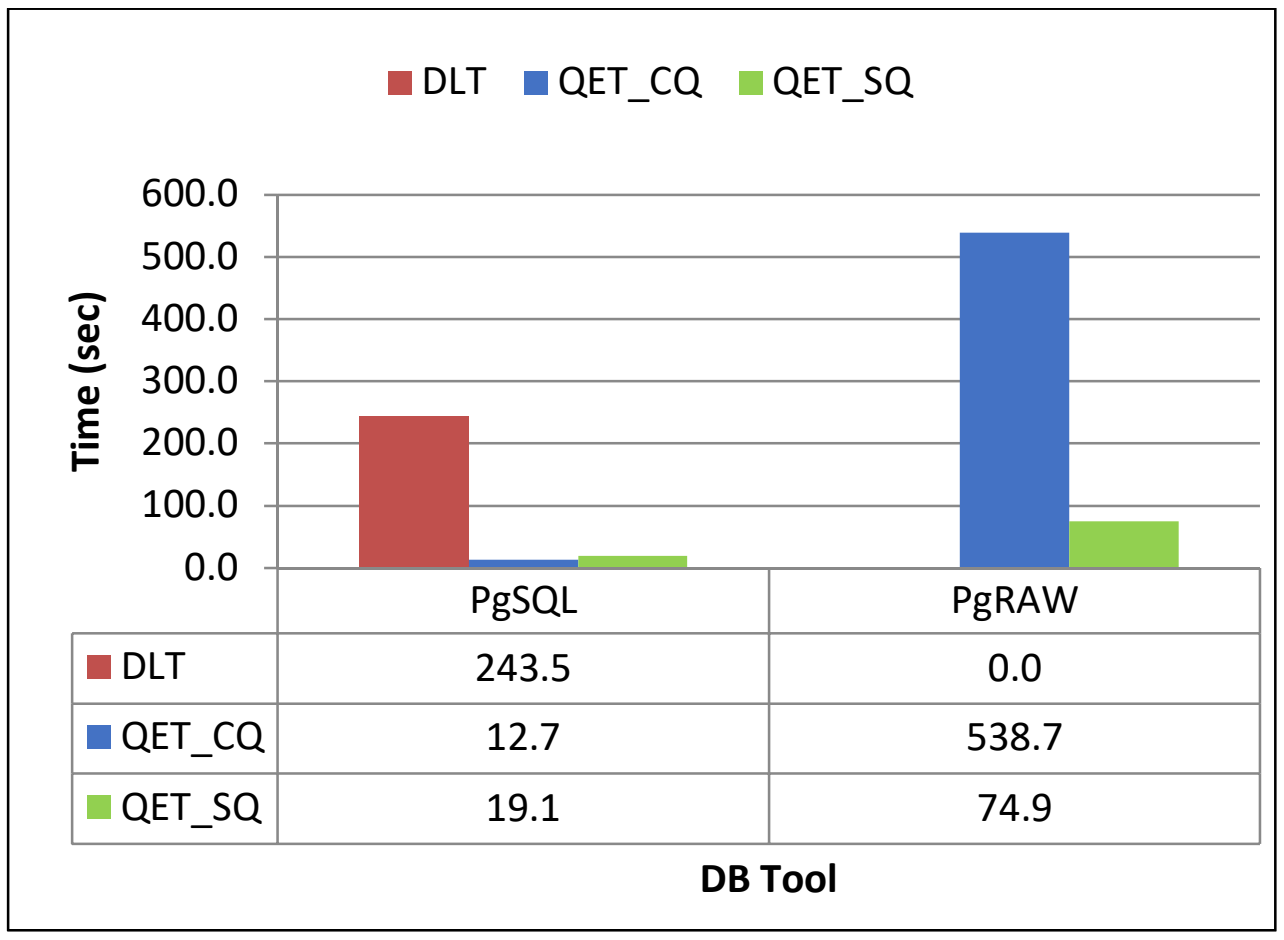

| | PgSQL | PgRAW |
|---|---|---|
| DLT | 243.5 | 0.0 |
| QET_CQ | 12.7 | 538.7 |
| QET_SQ | 19.1 | 74.9 |

Fig. 3. WET for SDSS Dataset

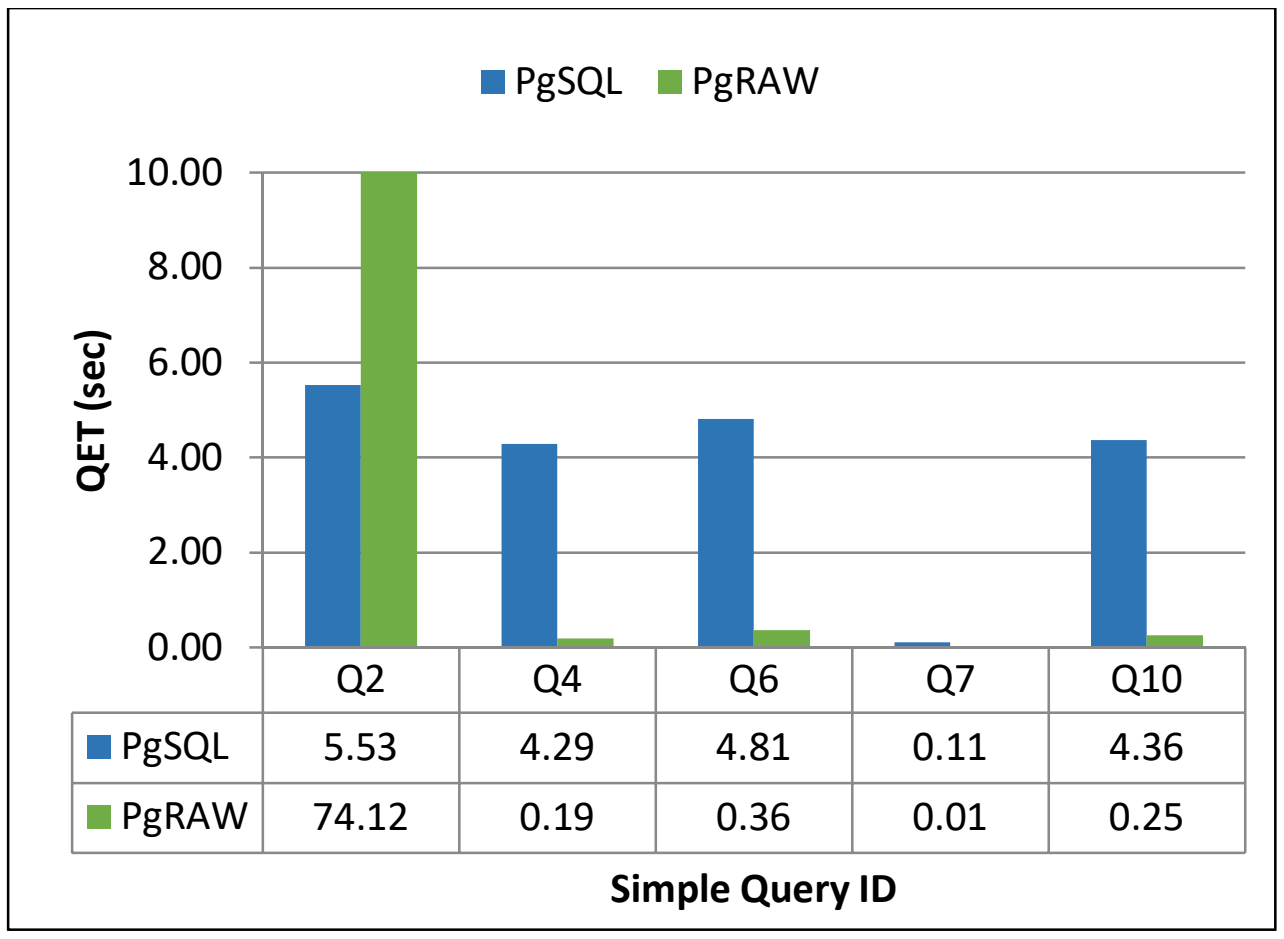

| | Q2 | Q4 | Q6 | Q7 | Q10 |
|---|---|---|---|---|---|
| PgSQL | 5.53 | 4.29 | 4.81 | 0.11 | 4.36 |
| PgRAW | 74.12 | 0.19 | 0.36 | 0.01 | 0.25 |

Fig. 4. QET: Simple Queries

## B. *QCA Technique*

This section discusses experimental results for the different data distribution cases discussed earlier. The QCA results have also been compared with the best results of workload aware WA partitioning techniques WSAC [14] and Vertical partitioning technique by Zhao [13]. The WA case represents the best results of these state-of-the-art techniques when enough storage budget is available to load all required attributes.

### 1) *Single Core Execution*

Fig. 5 compares query complexity aware QCA partitioned dataset's workload execution time for cases I, II, and V with the original dataset WET and workload aware WA techniques WET. The PgRAW is the original dataset WET. It can be seen that PgRAW took 87.8% of time executing complex queries while simple query execution time was only 12.2%. The WA technique load only required 54 workload attributes, reducing WET by 94.6%. The QCA case-I load all 34 attributes required by complex queries in the database, which improves CQ QET by 98.7%. The CQ QET improvement of case-I is 2.5% compare to WA due to the smaller partition size, as shown in Fig 6. The SQ QET is higher than WA in case-I because attributes required by simple queries reside on the storage device until the first SQ accesses the raw partition. Case-II has the lowest DLT due to the fewer attributes getting loaded into the database. The *CAP* partition is residing in raw format, increasing QET time of complex queries. However, case-II achieved the lowest QET time for simple queries because complex queries had already cached and indexed the raw data before the execution of SQ started. The case-V represents the replication of *CAP* on both formats. All the data required by each workload query is kept in a single partition, which means no additional joins are needed. The CQ achieved the lowest possible QET while SQ had to access the data from raw, which increased the SQ time by almost 9.5x times compared to case-II. Case I, II, and V reduced overall WET by 93.53%, 94.63% & 94.64% compared to the original dataset WET of PgRAW [5].

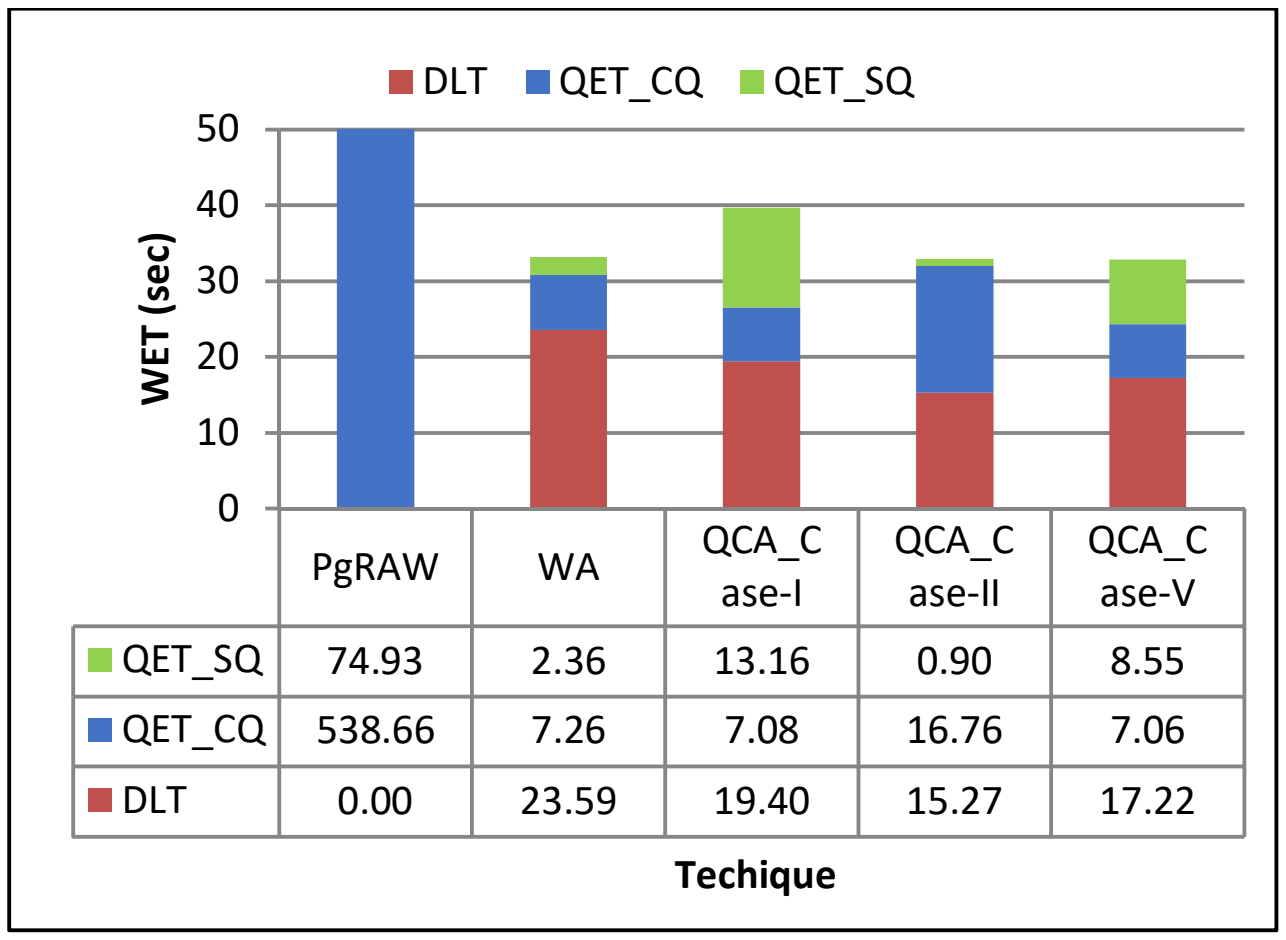

| | PgRAW | WA | QCA_Case-I | QCA_Case-II | QCA_Case-V |
|---|---|---|---|---|---|
| QET_SQ | 74.93 | 2.36 | 13.16 | 0.90 | 8.55 |
| QET_CQ | 538.66 | 7.26 | 7.08 | 16.76 | 7.06 |
| DLT | 0.00 | 23.59 | 19.40 | 15.27 | 17.22 |

Fig. 5. QCA-WET for SDSS Dataset

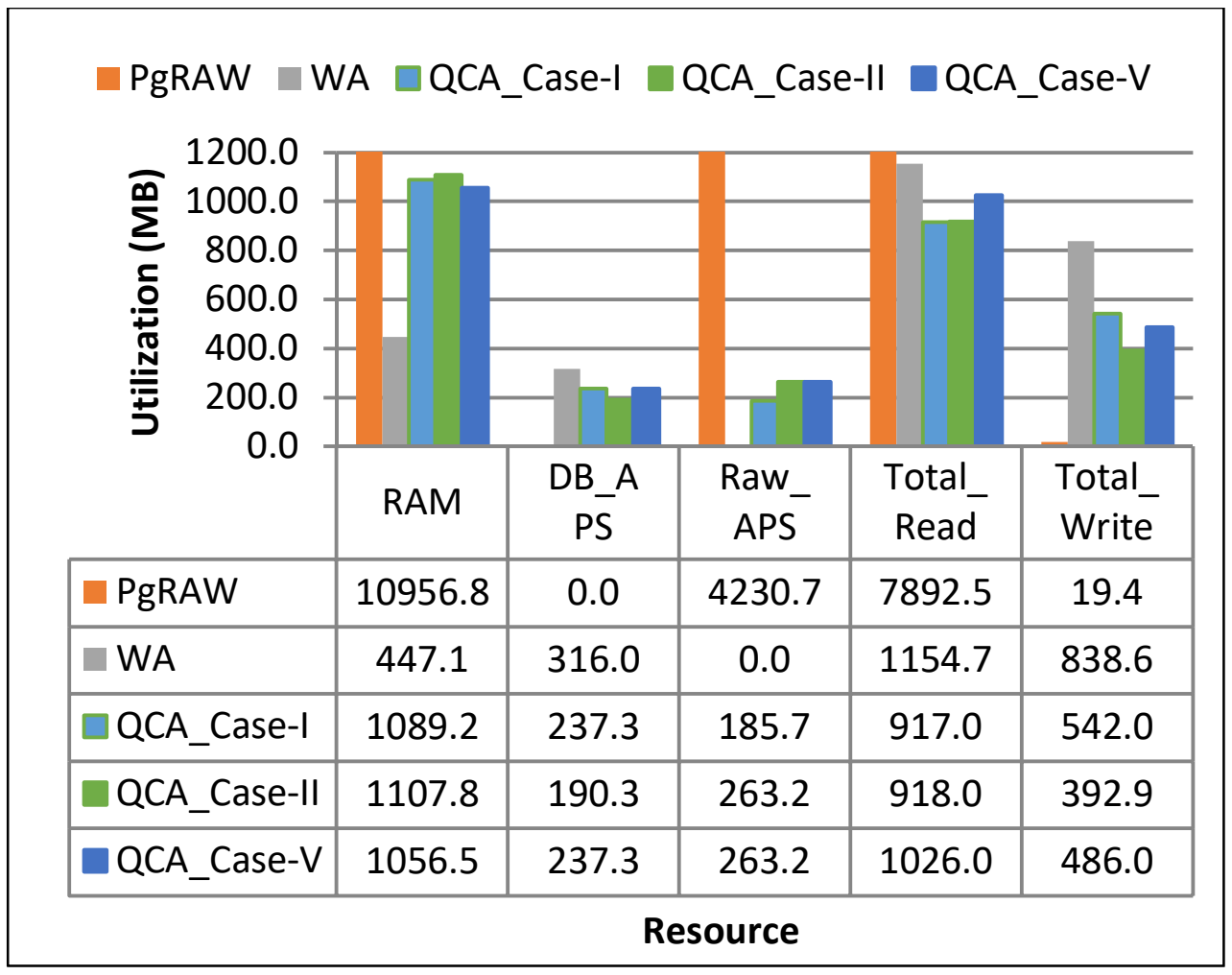

| | RAM | DB_APS | Raw_APS | Total_Read | Total_Write |
|---|---|---|---|---|---|
| PgRAW | 10956.8 | 0.0 | 4230.7 | 7892.5 | 19.4 |
| WA | 447.1 | 316.0 | 0.0 | 1154.7 | 838.6 |
| QCA_Case-I | 1089.2 | 237.3 | 185.7 | 917.0 | 542.0 |
| QCA_Case-II | 1107.8 | 190.3 | 263.2 | 918.0 | 392.9 |
| QCA_Case-V | 1056.5 | 237.3 | 263.2 | 1026.0 | 486.0 |

Fig. 6. Resource Utilization

Fig. 6 compares resources required by the PgRAW, WA techniques, and QCA cases. The RAM utilization is reduced by 89.9 – 95% by QCA and WA techniques. The QCA technique required 57-59% more RAM than WA because WA required a single DB tool. In comparison, QCA uses DB and Raw engine tools, which increases its RAM utilization. The accessed partition size APS of DB and Raw is displayed in the graph. All the APSs of WA & QCA cases are less than 7.5% in size compared to the original dataset size of 4.1GB. Smaller partition size reduces disk access, and required partitions occupy less RAM than the original non-partitioned dataset. The in-memory caching & indexing also become more manageable. The overall WET for WA, Case-II, and Case-V remained similar in single-core sequential workload execution. Therefore, WA and QCA techniques were applied to a multi-node and multi-core setup to identify their benefits and limitations.

*2) Optimizing Resource Utilization*

The parallel execution in a multi-node setup requires the data to be replicated on those nodes. The state-of-the-art data distribution techniques replicate the entire dataset or hot data partitions on multiple nodes and try to distribute query workload equally. Fig. 7 shows the comparison of WET when WA hot partition is distributed on multiple nodes N1 and N2 with the proposed QCA technique. It can be seen that WA requires hot data in loaded format on both nodes. While QCA only needs the CQ partition loaded on a single node. The WA can assign queries on hot partitions to balance the load on both nodes. In contrast, QCA distributes the CQ and SQ query workload to respective nodes. It can be observed that QCA distribution saved 15.9% WET or CPU time on Node 1 and 69.3% on Node 2 compared to WA. The average WET time saving is 42.66% on the presumed 2-node setup.

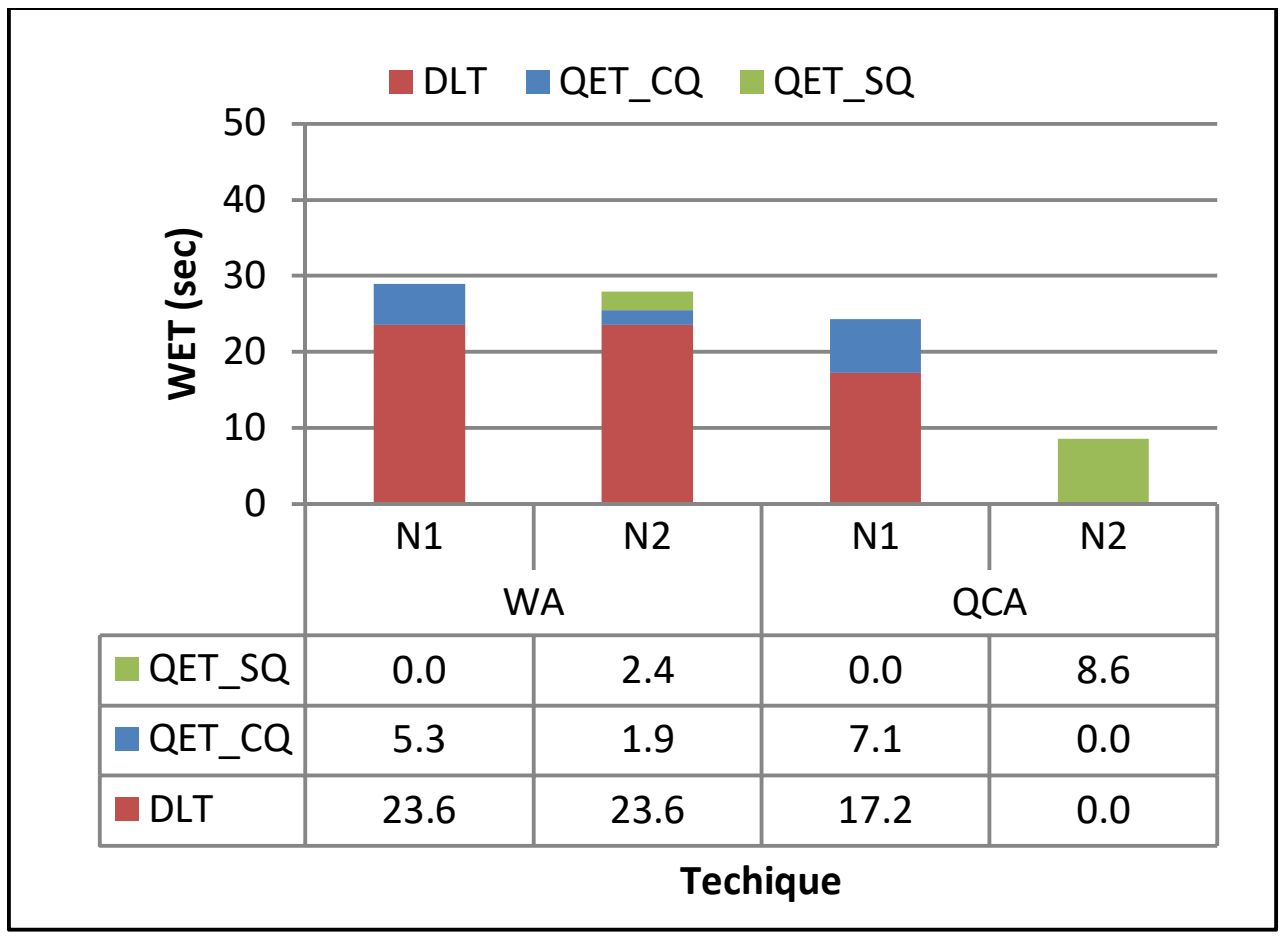

Fig. 7. Multi-node Execution

Researchers have observed that single-core workload execution does not utilize all available resources. In addition, data loading tasks cannot be executed in parallel to improve DLT for disk storage devices [20]. Most DBMS and WA case requires the data to be loaded into a database before query execution starts. The execution of queries cannot be started until the data loading completes, which needs data from database partitions. The parallel loading does not improve DLT [20]. Therefore, data loading task is performed sequentially. Multi-core CPU execution of SDSS workload for WA & QCA techniques has been compared in fig. 8 and 9. However, all workload queries can be executed in parallel as soon as data loading completes for WA. Fig. 8 shows the comparison of WET time for WA and QCA case-V. For QCA, data loading steps have been executed using a single thread on one CPU core, while SQ query execution is done in parallel. Fig.9 shows that for WA all CPU cores stay idle during the data loading task. While for QCA, core-1 is assigned a data loading task and core-2 executes the simple queries using SQ raw partition. The QCA completes SQ queries in parallel to CQ partition data loading as replicated *CAP* in SQ partition does not need any data from the database. The CQ queries were executed in parallel for both as soon as required partitions completed loading. It can be seen that QCA reduced the WET time by 25.46% compared to state-of-the-art WA techniques in a multi-core execution setup on a single node.

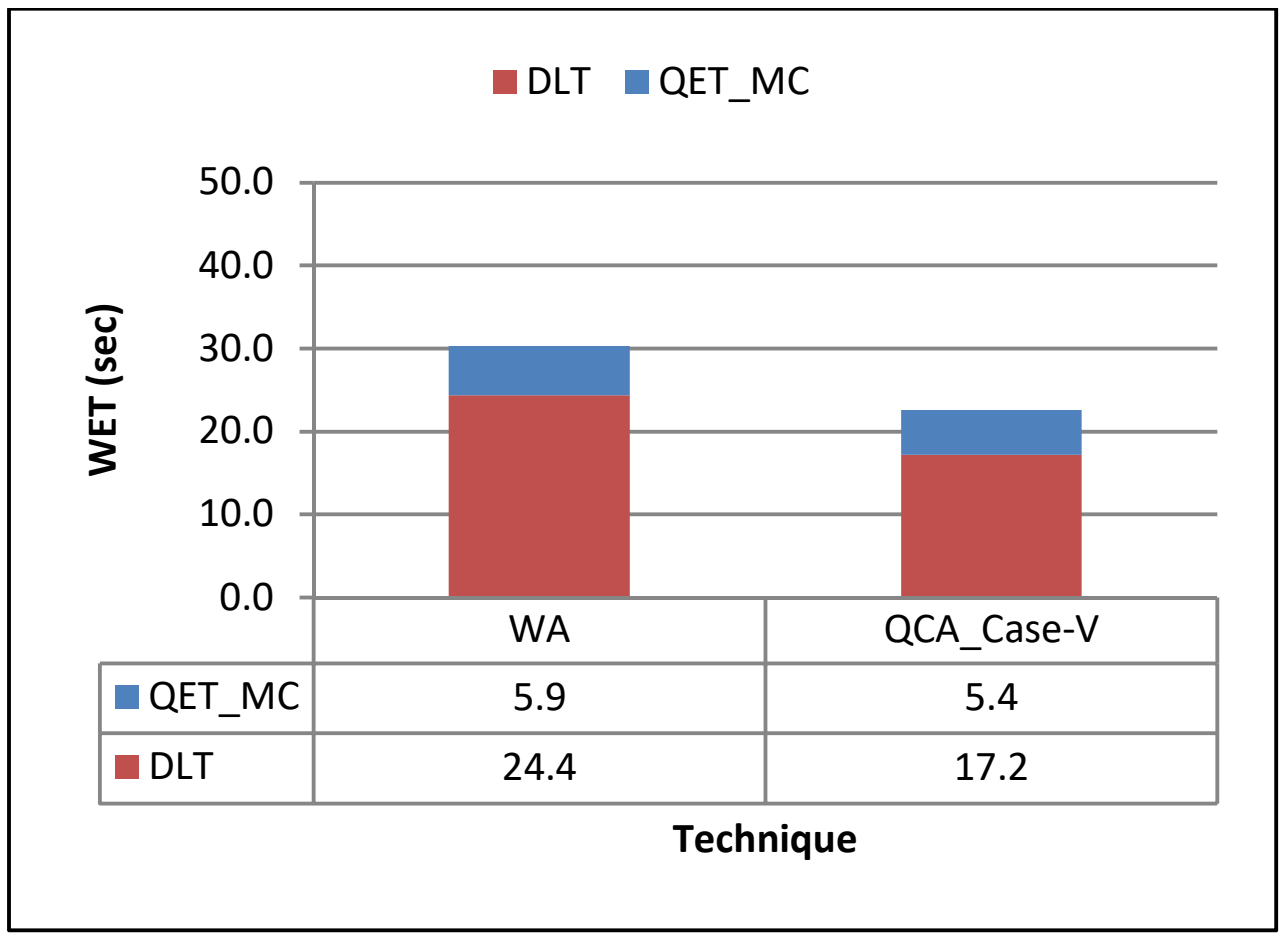

Fig. 8. Multi-core Execution on single node

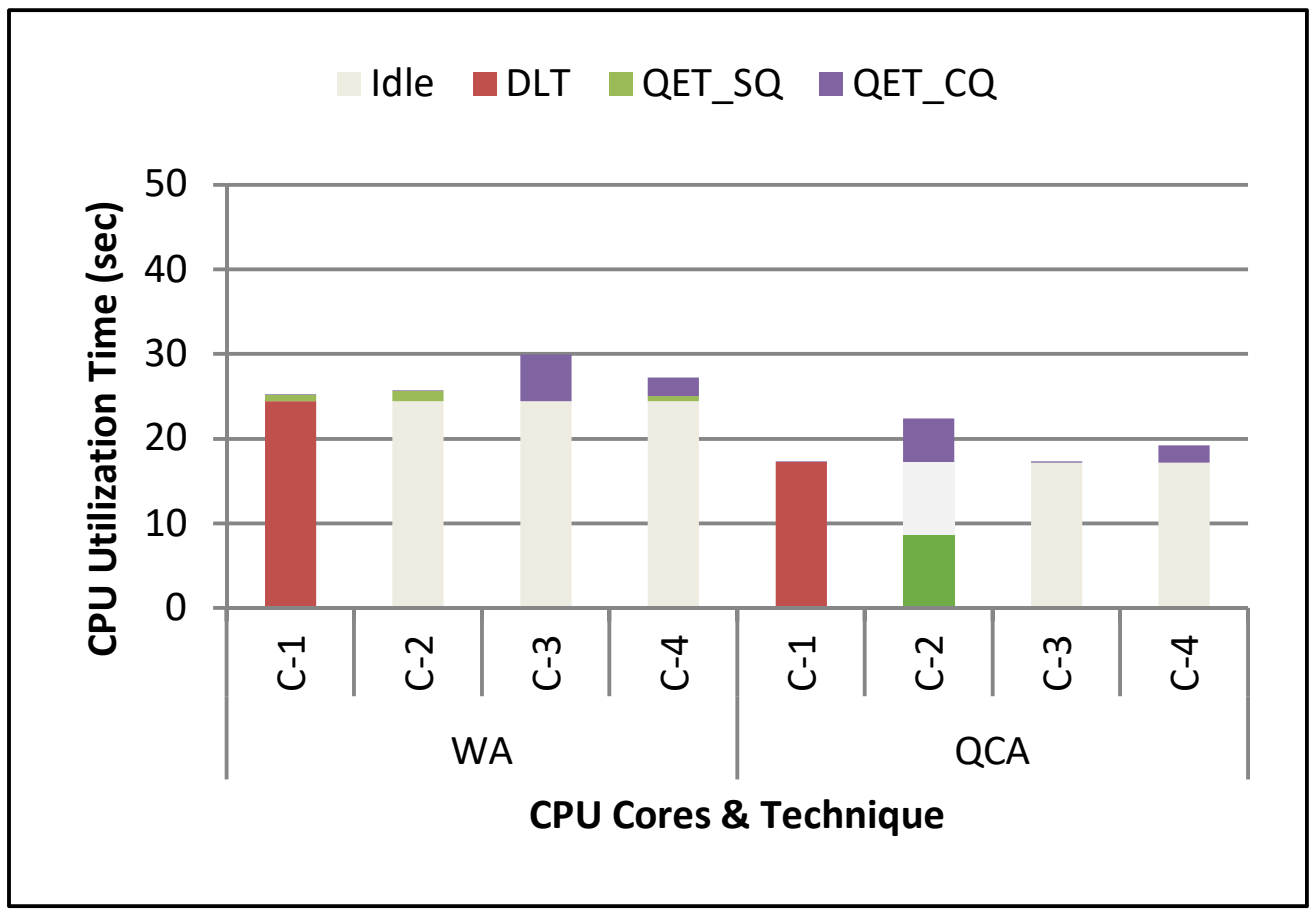

Fig. 9. Multi-core CPU Resource Optimization

## C. *Comparison with state-of-the-art techniques:*

Table III. shows a comparison of different data distribution techniques which use one or more nodes. The HTAP systems replicate 100% of data in multiple systems, as the used systems cannot perform joins on data existing in different formats. The 100% replication eliminated inter-node communication.

However entire dataset needs to be loaded on all nodes. The Zhao[13] and WSAC[14] techniques replicate data required by workload queries. The 89.4% of the SDSS dataset is not used by workload queries and is therefore not replicated. The PDC uses a horizontal partitioning technique to distribute large datasets among multiple nodes to achieve 0% replication. However, when a query requires data from other nodes, the required data or partial query results must be sent to a single node to produce the final result. The proposed QCA techniques achieve zero inter-node communication by replicating only 1.8% of the SDSS dataset for the extracted workload. The QCA technique does not require data loading on all nodes because the nodes processing simple queries cache processed data in RAM for faster query execution.

TABLE III. COMPARISON: PARTITIONING TECHNIQUES

| Technique | Partitioning | Replication % | Inter-node Comm. | Multi-Node Data Loading |
|---|---|---|---|---|
| HTAP [3] | - | 100% | No | Yes |
| Zhao [13] | VP | 10.6% | No | Yes |
| WSAC [14] | VP | 10.6% | No | Yes |
| PDC [18] | HP | 0% | Yes | Yes |
| QCA | VP | 1.8% | No | No |

## VI. CONCLUSION

The paper proposed a lightweight query complexity identification and partitioning technique QCA. The QCA identified two types of workload queries: zero join queries as simple queries SQ, and the remaining one or more join queries as complex queries CQ. The technique proposes partitioning the dataset in simple and complex query partitions to benefit from the observation that simple zero join queries execute faster with raw engines like PostgresRAW. The SQ partition is kept in raw format, while the CQ partition is loaded in the database. The paper discussed different data distribution cases on managing common attributes partition *CAP*. The result analysis of important data distribution cases I, II, and V is done in sequential and parallel execution setup. The QCA reduced the workload execution time WET of a real-world scientific dataset SDSS by 94.6% compared to the original dataset executed using the NoDB tool PostgresRAW. The single-node sequential execution of QCA could improve WET by 1.1% compared to workload aware WA techniques [13], [14]. However, parallel execution on a multi-node setup improved overall WET time by 42.66% compared to WA techniques. QCA increased CPU & RAM resources utilization during CQ partition loading to execute simple queries in parallel using SQ raw partition. A single node multi-core execution of QCA improved WET by 25.46% compared to the multi-core execution of WA. It can be concluded that QCA performed better than state-of-the-art workload ware WA techniques in a distributed environment while minimizing replication by 5.8x times.